# Anomalous attenuation of longitudinal ultrasound in intermediate state of high pure type I superconductor

A. G. Shepelev, O. P. Ledenyov and G. D. Filimonov

*National Scientific Centre Kharkov Institute of Physics and Technology, Academicheskaya 1, Kharkov 61108, Ukraine.*

The dependence of the ultrasonic attenuation on the direction of magnetic field $\Gamma(\varphi)$ in an intermediate state of very pure *Gallium* single crystal at the magnetic field $H$ ($H \perp k$) at the temperature $T = 0.5\,°K$ is found to be anomalously different from the dependence, observed at the same temperature and the magnitude of magnetic field equal to the critical magnetic field $H=H_c$ in a normal state. This phenomenon can be explained, considering the anisotropy of *Fermi* surface in the very pure *Gallium* single crystal.

## Introduction

In the process of experimental studies of high pure *Gallium* single crystal at the magnetic field $H$ at the temperature of $0.5°K$, we have found an anomalous change in the dependence of the attenuation of longitudinal ultrasound on the direction of magnetic field $\Gamma(\varphi)$ in an intermediate sate of high pure *type I* superconductor [1], comparing to the normal one.

## Measurements set up and methodology

The measurements were conducted in a $^3He$ cryostat [2]. The sample of extremely pure *Ga* (the single crystal was grown in vacuum by the *Chokhralsky* method) with the length / diameter ratio of *3* (the sample diameter of *7mm*) was investigated. The wave vector of sound $k$ was parallel to the *Ga* crystallographic axis $b$, oriented along the sample axis. The transducers were the plates of quartz of $X$-cut with thickness of *0.3 mm* and diameter of *4.5 mm*. Thus, the ultrasonic beam passed along the sample axis avoiding the regions near the side surfaces. The structure of intermediate state in the sample was formed with homogeneous transverse magnetic field $H$ ($H \perp k$) of *Helmholtz* pair. The magnetic field vector $H$ was able to rotate in the plane ($a$, $c$) of *Ga* crystallographic axes at a rate of *1* or *0.5 rev/min*, thus, leading to a structure of intermediate state close to an equilibrium and allowing one to record the dependence of the ultrasonic attenuation on the direction of magnetic field $H$. The stabilization accuracy of magnetic field was better than $3 \cdot 10^{-4}$. The measurements of ultrasonic attenuation could be done in the both modes by the automatic recording regime of the *PDS-021 x-y* recorder and by the points, using the relative method, i.e. by comparison of the signal under study with a standard pulse [3]. In the process of measurement, the sample was in direct contact with the liquid $^3He$, and the earth magnetic field was thoroughly compensated by the two pairs of *Helmholtz* coils. The researched *Ga* single crystal has the three important physical features:

(1) The *Pippard* oscillations in a normal state are observed in the high *Ga* single crystal at the very low magnetic fields in Fig. 1; thus, the electron free path is $l_e \sim 1\,cm$ approximately.

(2) The magnitude of ultrasonic attenuation is fully reversible at the transition of the sample from the superconducting to normal state under the influence of magnetic field and vice versa, i.e. a "frozen in" flux does not exist.

(3) In this high pure *Ga* single crystal, we have earlier found the giant oscillations of ultrasonic attenuation in an intermediate state of superconductor [4] – a phenomenon, theoretically predicted by *Andreev* [5].

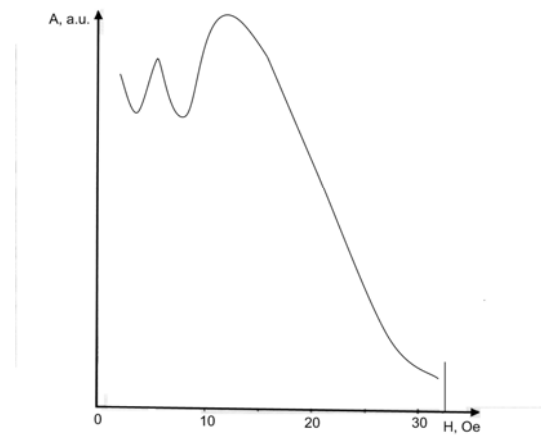

*Fig. 1. Amplitude dependence of 30 MHz longitudinal ultrasound transmitted through the Ga single crystal in normal state on magnetic field $H$ at $T= 2\,°K$, $k \parallel b$, $H \perp k$, $\angle H, c = 0°$.*

## Experimental measurements results

In this report, we give some results of investigation for the attenuation $\Gamma$ of *30 MHz* longitudinal ultrasound in the normal and intermediate states in high pure *Ga* single crystal. We have studied the dependence of $\Gamma$ upon the direction of magnetic field **H** in (**a**, **c**) plane of *Ga* crystallographic axes at a constant temperature of *0.5 °K*. Since, in the experiment, the condition of "high field" $l_e >> D$ is fulfilled (*D* is the diameter of electron's external orbit), and since the magnetic field exerts a deciding influence upon the electron dynamics, the ultrasonic attenuation in a *normal state* at $H = H_c$ depends appreciably upon the magnetic field orientation **H** in Fig. 2 – the range of $\Gamma$-change is $\simeq 6dB/sample$. The shape of this "rotational diagram" of dependence $\Gamma(\varphi)$ is practically symmetric, and connected with the anisotropy of *Fermi* surface in the high pure *Gallium* single crystal [6].

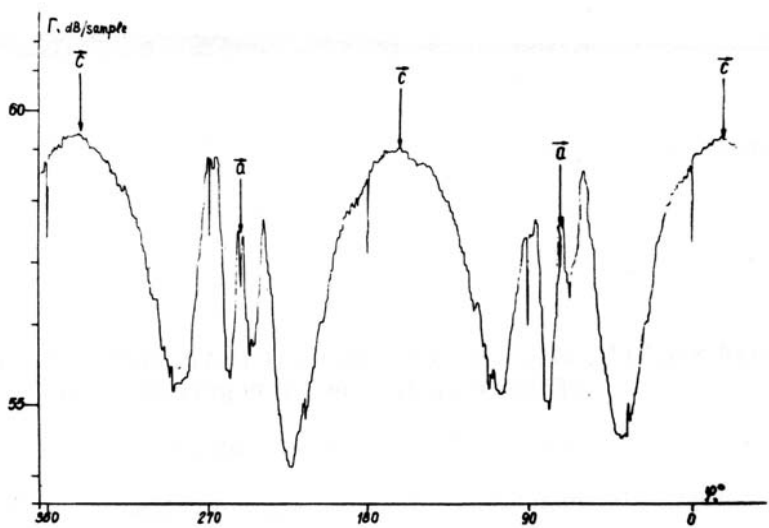

*Fig. 2. Attenuation dependence of 30 MHz longitudinal ultrasound on magnetic field direction in (**a**, **c**) crystallographic plane of Ga single crystal in normal state at T = 0.5 °K, H = 44 Oe, **k** ∥ **b**, **H** ⊥ **k**; $\varphi$ – angle values on limb of **H** rotation system, which automatically marks with single spikes angles multiple to 90°.*

In Fig. 3, the "rotation diagram" was obtained in the same experiment, but in the *Ga intermediate state* at slightly lowered magnitude of external magnetic field, corresponding to the normal phase concentration $c_n=0.9$. The intermediate state structure was formed by the rotation of **H** with an interval between the points; in which rotation was performed, $\Delta H=0.0025H_c$; the rate of decrease of **H** between the points was *0.7 Oe/min*. While the varying magnitudes of **H**, in the intermediate state oscillations of the $\Gamma$, were observed as in reference [4]. The curve of Fig. 3 strikingly differs from the curve of Fig. 2 both by its form and by the range of variation $\Gamma \simeq 12dB/sample$. The transition of high pure *type I* superconductor from a normal state to an intermediate state [1] at the magnetic field decrease is known to be followed by an appearance of a system with the alternating layers of the normal and superconducting phases; and the concentration of normal phase in the sample naturally decreases. *While the magnetic field is absent in the superconducting regions; it is equal to $H_c$ and directed along the layers in the regions, occupied by the normal phase in the rage of an intermediate state.* Since at the low temperature, the ultrasonic attenuation in superconducting phase is practically absent [7] and mainly occurs on the electrons in the layers with the normal phase during the transition from the normal to intermediate state, one might expect a decrease of $\Gamma$ proportional to $c_n$ in the sample. The form of dependence $\Gamma(\varphi)$, inherent to the normal state at $H=H_c$ in Fig. 2, would remain unchanged during this transition of the sample. However, these simple considerations can not explain the shape of the curve in Fig. 3.

In our opinion, the nature of the discovered phenomenon can be explained, using the *Andreev* theory [5], in spite of the fact that this theory is concerned with a model of metal with the simplest *isotropic Fermi* surface. The fact is that, in an intermediate state of superconductor, when the electrons encounter some peculiar reflections [8] on the inter-phase boundaries in the normal phase layers, and at the condition $l_e >> D$, the ultrasonic attenuation depends, in an oscillatory way, on the ratio $D/a(H)$, where *a* is the thickness of the normal phase layer. We observed the giant oscillations of ultrasound attenuation [4], which, in accordance with the *Andreev* theory [5], were caused by the $a/(H)$ change, characterized by the change of magnitude, but not the orientation of external magnetic field **H**, while the *D* was constant, since the magnetic field does not change and equals to the magnitude of critical magnetic field $H_c$ in the regions, occupied by the normal phase in an intermediate state of high pure *type I* superconductor.

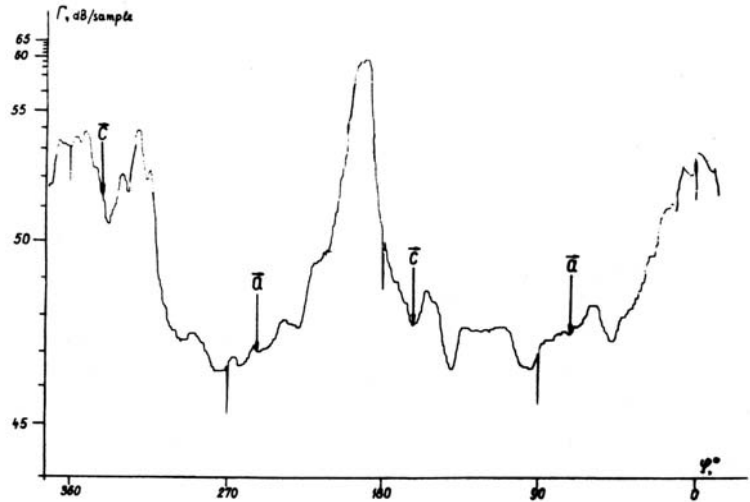

*Fig. 3. Attenuation dependence of 30 MHz longitudinal ultrasound on magnetic field direction in (**a, c**) crystallographic plane of Ga single crystal in intermediate state at T = 0.5 °K, **k** ∥ **b**, **H** ⊥ **k**, $c_n$ = 0.9; φ – angle values on limb of **H** rotation system, which automatically marks with single spikes angles multiple to 90°.*

In the present work, the orientation of magnetic field **H** was changed, but not its magnitude, hence, in the normal phase layers, the diameter of electron's external orbit $D$ was undergoing a change, because of anisotropy of *Ga Fermi* surface, while the thickness of normal phase layers $a$ and the magnitude of magnetic field $H=H_c$ within them were constant. In Fig. 3, some asymmetry of the curve results from the fact that the structure of an intermediate state in the considered region of normal phase concentrations is known to be not purely ideal. Therefore, the attenuation magnitude and the "rotation diagram" shape $\Gamma(\varphi)$ must be affected by its slight non-ideality as well.

The detected anisotropy of the "rotation diagram" $\Gamma(\varphi)$ in an intermediate state of high pure *type I* superconductor substantially increases at the increase of frequency of ultrasonic signal. In the magnetic field $H<H_c/2$, when the sample is in the superconducting state and the magnetic field **H** does not penetrate into it, the dependence of the ultrasound attenuation on the orientation of magnetic field $\Gamma(\varphi)$ is not observed. We decided to make some additional measurements at various concentrations of normal phase in an intermediate state of high pure *Ga* single crystal in our next research.

## Conclusion

The research on the attenuation of longitudinal ultrasound in an intermediate state of high pure *type I* superconductor is completed. The dependence of the ultrasonic attenuation on the direction of magnetic field $\Gamma(\varphi)$ in an intermediate state of very pure *Gallium* single crystal at the magnetic field **H** (**H** ⊥ **k**) at the temperature $T = 0.5$ °K is found to be anomalously different from the dependence, observed at the same temperature and the magnitude of magnetic field equal to the critical magnetic field $H=H_c$ in a normal state of superconductor. The theoretical explanation of observed phenomena is proposed, considering the anisotropy of *Fermi* surface in the very pure *Gallium* single crystal. It should be noted that the discovered phenomena can only appear in the high pure *type I* superconductors with the anisotropic *Fermi* surface.

The authors are grateful to Boris G. Lazarev for his interest in these experimental studies, to Alexander F. Andreev, E. A. Kaner, Yu. V. Sharvin for the thoughtful theoretical discussions, and to A. I. Berdovskii, V. V. Zhuk for their kind helps in the advanced measurements setup creation and high precision measurements completion.

This research paper was published in the *Solid State Communications* in 1973 [9].

*E-mail:   ledenyov@kipt.kharkov.ua

———————

1. Landau L. D. *Zh. Eksp. i Teor. Fiz.* **7**, 371 (1937).
2. Shepelev A. G. *Pribory i Tekn. Eksp.* no. 5, 237 (1971).
3. Shepelev A. G. and Filimonov G. D. *Cryogenics* **6**, 103 (1966).
4. Shepelev A. G., Ledenyov O. P., and Filimonov G. D. *Zh. Eksp. i Teor. Fiz. Pis. Red.* **14**, 428 (1971).
5. Andreev A. F. *Zh. Eksp. i Teor. Fiz.* **53**, 680 (1967).
6. Reed W. A. *Phys. Rev.* **188**, 1184 (1969).
7. Shepelev A. G. *Usp. Fiz. Nauk* **96**, 217 (1968).
8. Andreev A. F. *Zh. Eksp. i Teor. Fiz.* **46**, 1823 (1964).
9. A. G. Shepelev, O. P. Ledenyov, G. D. Filimonov, Anomalous attenuation of longitudinal ultrasound in the intermediate state of pure type I superconductor, *Solid State Communications*, vol. **12**, issue 3, pp. 241-244, ISSN: 0038-1098, 1973; http://www.sciencedirect.com/science/article/pii/003810 9873905103 .